\documentclass[longauth,structabstract]{aa}
\usepackage{natbib}
\usepackage{graphicx}
\usepackage{txfonts}
\usepackage{multirow}
\begin{document}
%
%\preprint{MPP-2009-224, \tt arXiv:1001.1291v2 [astro-ph.CO]}
\title{MAGIC TeV Gamma-Ray Observations of Markarian~421\\during Multiwavelength Campaigns in 2006}
\titlerunning {Observations of Mkn\,421 during MWL Campaigns in 2006}
\authorrunning{MAGIC Collaboration}
\offprints{{\tt snruegam@astro.uni-wuerzburg.de}, {\tt oya@gae.ucm.es}, {\tt robert.wagner@mpp.mpg.de}}

\author{
 J.~Aleksi\'c\inst{1} \and
 H.~Anderhub\inst{2} \and
 L.~A.~Antonelli\inst{3} \and
 P.~Antoranz\inst{4} \and
 M.~Backes\inst{5} \and
 C.~Baixeras\inst{6} \and
 S.~Balestra\inst{4} \and
 J.~A.~Barrio\inst{4} \and
 D.~Bastieri\inst{7} \and
 J.~Becerra Gonz\'alez\inst{8}
% ...
}
\institute { IFAE, Edifici Cn., Campus UAB, E-08193 Bellaterra, Spain
 \and ETH Zurich, CH-8093 Switzerland
 \and INAF National Institute for Astrophysics, I-00136 Rome, Italy
 \and Universidad Complutense, E-28040 Madrid, Spain
 \and Technische Universit\"at Dortmund, D-44221 Dortmund, Germany
 \and Universitat Aut\`onoma de Barcelona, E-08193 Bellaterra, Spain
 \and Universit\`a di Padova and INFN, I-35131 Padova, Italy
 \and Inst. de Astrof\'{\i}sica de Canarias, E-38200 La Laguna, Tenerife, Spain
% ...
}

\date{Received 22 December 2009 / Accepted 19 May 2010}

\abstract
% context heading (optional)
{Wide-range spectral coverage of blazar-type active galactic
nuclei is of paramount importance for understanding the particle acceleration
mechanisms assumed to take place in their jets. The Major Atmospheric Gamma
Imaging Cerenkov (MAGIC) telescope participated in three multiwavelength (MWL)
campaigns, observing the blazar \object{Markarian (Mkn) 421} during the nights
of April 28 and 29, 2006, and June 14, 2006.}
% aims heading (mandatory)
{We analyzed the corresponding MAGIC very-high energy observations during 9
nights from April 22 to 30, 2006 and on June 14, 2006. We inferred light curves
with sub-day resolution and night-by-night energy spectra.}
% methods heading (mandatory)
{MAGIC detects $\gamma$-rays by observing extended air showers in the
atmosphere. The obtained air-shower images were analyzed using the standard MAGIC
analysis chain.}
% results heading (mandatory)
{A strong $\gamma$-ray signal was detected from \object{Mkn\,421} on all
observation nights. The flux ($E>250$~GeV) varied on night-by-night basis
between $(0.92\pm0.11)\,10^{-10} \mathrm{cm}^{-2} \mathrm{s}^{-1}$ (0.57 Crab
units) and $(3.21\pm0.15)\,10^{-10} \mathrm{cm}^{-2} \mathrm{s}^{-1}$ (2.0 Crab
units) in April 2006.  There is a clear indication for intra-night variability
with a doubling time of $36\pm10_{\mathrm stat}$ minutes on the night of 
April 29, 2006, establishing once more rapid flux variability for this object. For
all individual nights $\gamma$-ray spectra could be inferred, with power-law
indices ranging from 1.66 to 2.47. We did not find statistically significant
correlations between the spectral index and the flux state for individual
nights. During the June 2006 campaign, a flux substantially lower than the one
measured by the Whipple 10-m telescope four days later was found.  Using a
log-parabolic power law fit we deduced for some data sets the location of the
spectral peak in the very-high energy regime. Our results confirm the
indications of rising peak energy with increasing flux, as expected in leptonic
acceleration models.
}
{}

\keywords{Gamma rays: galaxies -- BL~Lacertae objects: individual (Mkn\,421) -- Radiation mechanisms: non-thermal}

\maketitle

\section{Introduction}
The active galactic nucleus (AGN) \object{Markarian (Mkn)\,421} was the first
extragalactic source detected in the TeV energy range, using imaging
atmospheric Cerenkov telescopes \citep[IACTs;][]{punch,petry}. With a redshift
of $z=0.030$ it is the closest known and, along with \object{Mkn~501}, the
best-studied TeV $\gamma$-ray emitting blazar.\footnote{See, e.g.,
http://www.mpp.mpg.de/$\sim$rwagner/sources/ for an up-to-date list of VHE
$\gamma$-ray sources.} So far, flux variations by more than one order of
magnitude \citep[e.g.,][]{fossati08}, and occasional flux doubling times as
short as 15~min \citep{gaidos,hegra421,swl08} have been observed. Variations in
the hardness of the TeV $\gamma$-ray spectrum during flares were reported by
several groups \cite[e.g.][]{whipple421,hess421,fossati08}. Simultaneous
observations in the X-ray and very-high energy (VHE; $E\gtrapprox
100~\mathrm{GeV}$) bands show strong evidence for correlated flux variability
\citep{krawczynski,blazejwski,fossati08}. With a long history of observations,
\object{Mkn\,421} is an ideal candidate for long-term and statistical studies
of its emission \citep{tskb07,g07,h09}.

\object{Mkn\,421} has been detected and studied at basically all wavelengths of
the electromagnetic spectrum from radio waves up to VHE $\gamma$-rays. Its
wide-range spectral energy distribution (SED) shows the typical double-peak
structure of AGN. \object{Mkn\,421} is a so-called blazar. These constitute a
rare subclass of AGNs with beamed emission closely aligned to our line of
sight.  In blazars, the low-energy peak at keV energies is thought to arise
dominantly from synchrotron emission of electrons, while the origin of the
high-energy (GeV-TeV) bump is still debated.  The SED is commonly interpreted
as being due to the beamed, non-thermal emission of synchrotron and
inverse-Compton radiation from ultrarelativistic electrons.  These are assumed
to be accelerated by shocks moving along the jets at relativistic bulk speed.
For most of the observations, the SED can be reasonably well described by
homogeneous one-zone synchrotron-self-Compton (SSC) models
\citep[e.g.][]{mg85,mgc92,costamante}. Hadronic models \citep{mannheim,muecke},
however, can also explain the observed features. A way to distinguish between
the different emission models is to determine the positions, evolution and
possible correlations \cite[see, e.g.,][for a review]{wagner08} of both peaks
in the SED, using simultaneous, time-resolved data covering a broad energy
range, e.g., as obtained in multiwavelength (MWL) observational campaigns.

In this Paper we present results from Major Atmospheric Gamma-ray Imaging Cerenkov
(MAGIC) telescope  VHE $\gamma$-ray observations of \object{Mkn\,421} during
eight nights from April 22 to 30, 2006, and on June 14, 2006. For most of the
days, optical $R$-band observations were conducted with the KVA telescope.
Simultaneous observations were performed by {\it Suzaku} \citep{suzaku} and
H.E.S.S., as well as by {\it XMM-Newton} \citep{xmm} on April 28 and 29, 2006,
respectively. During both nights, we carried out particularly long,
uninterrupted observations in the VHE energy band of $\approx 3$~hours duration
each.  An onset of activity in the X-ray band triggered an {\em INTEGRAL}-led
target-of-opportunity (ToO) campaign, which took place from June 14 -- 25, 2006
for a total of 829~ks \citep{lichtiaa}. Within this campaign, MAGIC observed
\object{Mkn\,421} at rather high zenith angles from 43 to 52 degrees in
parallel with {\em INTEGRAL} on June 14, 2006.

In the following sections, we describe the data sets and the analysis applied
to the VHE $\gamma$-ray data, the determination of spectra for all observation
nights, and put the results into perspective with other VHE $\gamma$-ray
observations of \object{Mkn\,421}.  The interpretation of these data in a MWL
context is presented in \citet{acci421} and subsequent papers.

VHE $\gamma$-ray observations in April and June 2006 have also been carried out
by the Whipple telescope \citep{horan}, by the VERITAS \citep{fegan07}, and
TACTIC \citep{yadav07} collaborations, although not simultaneously with our
observations.

\section{The MAGIC telescope}
The VHE $\gamma$-ray observations were conducted with the MAGIC telescope
located on the Canary island La~Palma (2200 m above sea level,
28$^\circ$45$'$N, 17$^\circ$54$'$W). At the time of our observations in 2006,
MAGIC was a single-dish 17-m~\O\ instrument\footnote{Since 2009, MAGIC is a
two-telescope stereoscopic system \citep{corti09}.} for the detection of
atmospheric air showers induced by $\gamma$-rays. Its hexagonally-shaped camera
with a field of view (FOV) of $\approx3.5^\circ$ mean diameter comprises 576
high-sensitivity photomultiplier tubes (PMTs): 180 pixels of $0.2^\circ$~\O\
surround the inner section of the camera of 394 pixels of $0.1^\circ$~\O\ ($=
2.2^\circ$~\O\ FOV). The trigger is formed by a coincidence of $\geq 4$
neighboring pixels. Presently the accessible trigger energy range (using the
MAGIC standard trigger; \citealt{trigger}) spans from $50-60$~GeV (at small
zenith angles) up to tens of TeV.
Further details, telescope parameters, and performance information can be found
in \citet{baixeras04,cortina,magic_cn}.

\section{Observations and data analysis}\label{observations}
The observations were carried out during dark nights, employing the so-called
wobble mode \citep{daum}, in which two opposite sky directions, each
0.4$^\circ$ off the source, are tracked alternatingly for 20 minutes each. 
The on-source data are defined by calculating image parameters with respect to the source
position, whereas background control (``off'') data are obtained from
the same data set, but with image parameters calculated with respect to
three positions, arranged symmetrically to the on-source region with
respect to the camera center.
The simultaneous measurement of signal and background makes additional
background control data unnecessary. In order to avoid an unwanted contribution
from source $\gamma$-events in the off sample, and to guarantee the statistical
independence between the on and the off samples in the signal region, events
included in the signal region of the on sample were excluded from the off
sample and vice versa.

The data were analyzed following the standard MAGIC analysis procedure
\citep{bw03,b08}. After calibration \citep{MAGIC_calibration} and extracting
the signal at the pulse maximum using a spline method, the air-shower images
were cleaned of noise from night-sky background light by applying a three-stage
image cleaning. The first stage requires a minimum number of 6 photoelectrons
in the core pixels and 3 photoelectrons in the boundary pixels of the images
\citep[see, e.g.][]{fegan97}.  These tail cuts are scaled according to the
larger size of the outer pixels of the MAGIC camera.  Only pixels with at least
two adjacent pixels with a signal arrival time difference lower than 1.75~ns
survive the second cleaning stage.  The third stage repeats the cleaning of the
second stage, but requires only one adjacent pixel within the 1.75~ns time
window.

The data were filtered by rejecting trivial background events, such as
accidental noise triggers, triggers from nearby muons, or data taken during
adverse atmospheric conditions (e.g., low atmospheric transmission).
12.7~hours out of the total 15.0~hours' worth of data survived the latter
quality selection and were used for further analysis.

We calculated image parameters \citep{h85} such as WIDTH, LENGTH, SIZE, CONC,
M3LONG (the third moment of the light distribution along the major image axis),
and LEAKAGE (the fraction of light contained in the outermost ring of camera
pixels) for the surviving events.  For the $\gamma$/hadron separation, a
SIZE-dependent parabolic cut in
AREA$\,\equiv\,$WIDTH~$\times$~LENGTH~$\times~\pi$ was used \citep{riegel05}.
The cut parameters for the assessment of the detection significance were
optimized on \object{Mkn\,421} data from close-by days. For the data of 
June 14, 2006 at rather large zenith angles, data of \object{Mkn\,501} from
October 2006 were used to determine the optimal cuts. Any significance in this
work was calculated using Eq. 17 of \citet{LiMa} with $\alpha=1/3$.

The primary $\gamma$-ray energies were reconstructed from the image parameters
using a Random Forest regression method \citep[][and references
therein]{magic_rf} trained with Monte-Carlo simulated events
\citep[MCs;][]{corsika,majumdar}. The MC sample is characterized by a power-law
spectrum between 10~GeV and 30~TeV with a differential spectral photon index of
$\alpha=-2.6$, and a point-spread function resembling the experimental one. The
events were selected to cover the same zenith distance range as the data.  For
the spectrum calculation, the area cut parameters were optimized to yield a
constant MC cut efficiency of 90\% over the whole energy range, increasing the
$\gamma$-ray event statistics at the threshold.  

The \object{Mkn\,421} observations presented here are among the first data
taken by MAGIC after major hardware updates in April 2006 \citep{mux}, which
required us to thoroughly examine the data. Despite the hardware changes, the
MAGIC subsystems performed as expected with the exception of an unstable
trigger behavior for some PMTs, leading to a significant loss of events in one
of the six sectors of the camera. In order to proceed with the data analysis
with serenity and to estimate the effect caused by this inhomogeneity, a simple
procedure was applied to the data: The expected number of events, as a function
of energy, for the affected sector was estimated as the mean of the number of
events in the other five sectors of the camera. (A homogeneous distribution of
events through the six sectors is expected for normal conditions). 
The difference between the expected and actually measured events was computed
using the whole data sample in order to have sufficient statistics. We
found a decrease of the differential photon flux of 5.7\% between 250 and
400~GeV, 4.6\% between 400 and 650~GeV, 2.2\% between 650 and 1050~GeV and $<
1$\% for higher energies for the April 2006 data. Due to the higher zenith
distance and energy threshold, the method was adapted for June 14, 2006 and
yielded a decrease of 5.2\% between 450 and 670~GeV and 2.6\% for higher
energies.  However the above mentioned effect is just an average one, with
estimated flux errors of up to 6.6\% showing up for individual nights.

To mitigate the effect of the inhomogeneity, instead of an (already increased)
energy threshold of 250~GeV, higher thresholds of 350 or 450~GeV were applied
for some observation nights. In this way we made sure that the estimated
systematic error remains within reasonable limits.

For the calculation of the individual light curves as well as for the overall
April 2006 lightcurve, the flux between 250~GeV and 350~GeV was extrapolated
for the nights with higher threshold. We assumed a power-law behavior in this
energy range, with the spectral index determined for the first three energy bins of
the whole April dataset (i.e.,\ $\alpha=-2.08$). The flux normalization for
each night has been determined at 500~GeV by a fit to the first three
differential spectral points, an energy range which is reliable for all
affected nights.

Tab.~\ref{tab:data} summarizes the analyzed data sets. The statistical
significance of any detection is assessed by applying a cut in $\theta^{2}$,
where $\theta$ is the angular distance between the expected source position and
the reconstructed $\gamma$-ray arrival direction. The arrival directions of the
showers in equatorial coordinates were calculated using the DISP method
\citep{fomin,lessard}.  We replaced the constant coefficient $\xi$ in the
parameterization of DISP in the original approach by a term which is dependent
on LEAKAGE, SIZE, and SLOPE,
\begin{equation}
\xi = \xi_0 + \xi_1 \, {\rm SLOPE} + \xi_2 \, {\rm LEAKAGE} + k \xi_3
\, (\log_{10}{\rm SIZE} - \xi_4)^2,
\end{equation}
$k = 0$ for $\log_{10}{\rm SIZE} < \xi_4$ and $k = 1$ for $\log_{10}{\rm SIZE}
\geq \xi_4$. The coefficients were determined using simulated data.  The
parameter SLOPE is a measure for the longitudinal arrival time evolution of the
shower in the camera plane similar to the time parameter GRADIENT in
\citet{mtime}. Instead of defining the parameter from a fit to the arrival time
distribution, however, SLOPE is determined as an analytical solution of the
fit. Note that this new parametrization makes DISP and therefore $\theta^2$
source dependent.

\begin{table}
\caption{\label{tab:data} Some characteristic parameters of the different data sets of the campaign.}
\begin{tabular}{cccc}
\hline Night & Observation Window [MJD] & $t_\mathrm{eff.}$ [h] &ZA [$^{\circ}$] \\
\hline \hline
April 22, 2006 & $53847.97679 - 53848.01460$ & 0.76 & 18 -- 28 \\
April 24, 2006 & $53849.96428 - 53850.00669$ & 0.99 & 16 -- 28 \\
April 25, 2006 & $53850.92813 - 53850.99607$ & 1.53 & 10 -- 26 \\
April 26, 2006 & $53851.92862 - 53852.00383$ & 1.64 & 10 -- 29 \\
April 27, 2006 & $53852.93474 - 53853.00047$ & 1.42 & 12 -- 28 \\
April 28, 2006 & $53853.88173 - 53854.01394$ & 2.23 & 10 -- 32 \\
April 29, 2006 & $53854.89514 - 53855.04119$ & 2.78 & \phantom{0}9 -- 41 \\
April 30, 2006 & $53855.97283 - 53855.97906$ & 0.16 & 23 -- 24 \\
June 14, 2006 & $53900.91979 - 53900.95532$ & 0.80 & 43 -- 52 \\
\hline
\end{tabular}

{\bf Notes.} $t_\mathrm{eff.}$ denotes the effective observation time. ZA gives the zenith angle range of the observations.
\end{table}

All stated errors are statistical errors only; we estimate our systematic
errors to be 16\% for the energy scale, 11\% for absolute fluxes and flux
normalizations, and $0.2$ for the spectral slopes \citep{magic_cn}, not
including the additional systematic flux errors mentioned above.

A second, independent analysis of the data yielded compatible results to those
presented here.

\section{Results}
\subsection{Results for April 22 -- 30, 2006}

\begin{figure*}
\includegraphics[width=\linewidth]{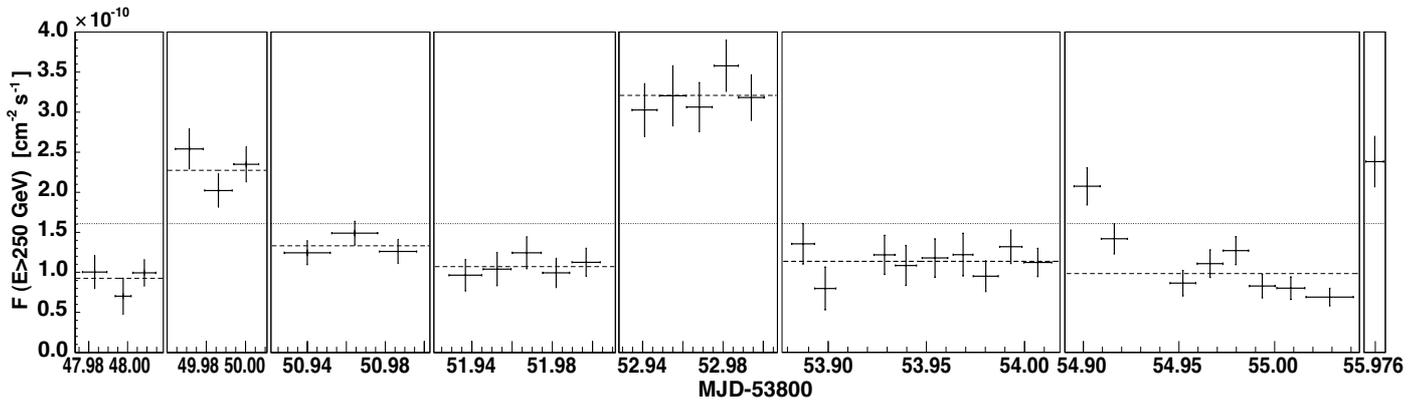}
\caption{VHE ($E>250\,\mathrm{GeV}$) light curve for
\object{Mkn\,421} observations in April 2006. The dotted line represents the
\object{Crab nebula} flux \citep{magic_cn}, whereas the individual dashed lines
show the result of a fit to the time bins (average nightly flux) of the
corresponding nights.} \label{fig:lcapril06fine}
\end{figure*}

\begin{figure}
\includegraphics[width=\linewidth]{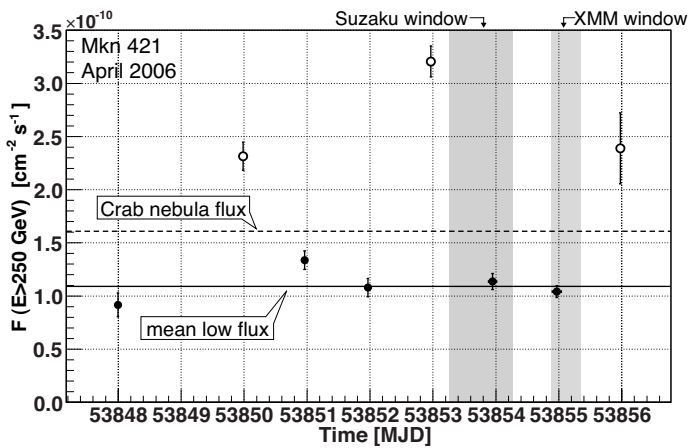}
\caption{VHE ($E>250$~GeV) light curve for
\object{Mkn\,421} observations in April 2006. The data points represent average
nightly fluxes. The observation windows of the {\it Suzaku} (MJD
53853.28--53854.27) and {\it XMM-Newton} (MJD 53854.87--53855.35) MWL campaigns
are marked by the gray-shaded areas. A ``mean low flux'' (solid line) was
averaged over all data points below $1.6\cdot 10^{-10} \mathrm{cm}^{-2}
\mathrm{s}^{-1}$, i.e., excluding those data points marked by thin open
circles.  The dashed line gives the \object{Crab nebula} flux \citep{magic_cn}
for comparison.} \label{fig:lcapril06}
\end{figure}

MAGIC observed \object{Mkn\,421} from MJD 53847 to MJD 53855.  During the
observations, two MWL campaigns were carried out simultaneously
with {\it Suzaku} and with {\it XMM-Newton} on MJD 53854 and MJD 53855,
respectively. \object{Mkn\,421} was also observed as part of the  monitoring
program of the Whipple 10-m telescope \citep[see][]{horan}, albeit about 3.5
hours after the MAGIC observations stopped, due to the different
longitudes of the two instruments.

A strong $\gamma$-ray signal from the source was detected in all eight
observation nights. In total, 3165 excess events were recorded over a
background of 693 events for energies $> 250$~GeV, yielding an overall significance of $64.8\sigma$.
\object{Mkn\,421} exhibited an average flux of
$F_{>250\,\mathrm{GeV}}=(1.48\pm0.03)\cdot10^{-10}\, \mathrm{cm}^{-2}
\mathrm{s}^{-1}$.  When compared to earlier observations \citep[see,
e.g.][]{magic421,tskb07,g07,steele}, our observations indicate an elevated flux
state of \object{Mkn\,421}.  We found high flux states in the nights of
MJD 53850, $F_{>250\,\mathrm{GeV}}=(2.32\pm0.13)\cdot10^{-10}\, \mathrm{cm}^{-2} \mathrm{s}^{-1}$,
MJD 53853, $F_{>250\,\mathrm{GeV}}=(3.21\pm0.15)\cdot10^{-10}\, \mathrm{cm}^{-2} \mathrm{s}^{-1}$, and
MJD 53856, $F_{>250\,\mathrm{GeV}}=(2.39\pm0.33)\cdot10^{-10}\, \mathrm{cm}^{-2} \mathrm{s}^{-1}$ (Fig.~\ref{fig:lcapril06}).
In the remaining nights (we assumed nights with fluxes below $1.6\cdot 10^{-10}
\mathrm{cm}^{-2} \mathrm{s}^{-1}$ as non-flare nights), Mkn 421 exhibited a
low-flux average of $F_{>250\,\mathrm{GeV}}=(1.09\pm 0.03)\cdot10^{-10}\,
\mathrm{cm}^{-2} \mathrm{s}^{-1}$.  The analysis results on a night-by-night
basis are summarized in Tab.~\ref{tab:aprilspectra}, and include the nightly
numbers for excess and background events, significances, and average integral fluxes
above $250$~GeV (where the nights with an energy cut of $350$~GeV where
extrapolated down to $250$~GeV, see Sect. \ref{observations} for details).
The results of a spectral fit based on a simple power law (PL) of the form
\begin{equation}
\frac{\mathrm{d}F}{\mathrm{d}E} = f_0 \cdot  10^{-10}\,
\mathrm{TeV}^{-1}\mathrm{cm}^{-2} \mathrm{s}^{-1} \,
\left(\frac{E}{E_0}\right)^{-\alpha}
\label{eq:2}
\end{equation}
\noindent are also shown.

\begin{table*}
\caption{\label{tab:aprilspectra} Analysis results.}
\begin{center}
\begin{tabular}{crrcccrccc}
\hline Observation Night & $N_\mathrm{excess}$ & $N_\mathrm{backgr.}$ & $S$ & $E_{\rm cut}$ [GeV] & $F(E>E_\mathrm{min})$ & $\chi^2_\mathrm{red,const}$ & $f_0$ & $\alpha$ & $\chi^2_\mathrm{red,PL}$ \\
\hline \hline
April 22, 2006 & 100 &  29 &                  10.9$\sigma$ & 350 & $0.92\pm0.11$ & \phantom{0}1.3/2 &                                              $0.98\pm0.15$ & $2.05\pm0.21$ & \phantom{0}2.1/2 \\
April 24, 2006 & 419 &  69 &                  25.0$\sigma$ & 250 & $2.32\pm0.13$ & \phantom{0}2.7/2 & \phantom{0} $2.45\pm0.14$ \phantom{0} & $2.25\pm0.09$ & \phantom{0}2.0/3 \\
April 25, 2006 & 342 &  83 &                  20.8$\sigma$ & 250 & $1.34\pm0.09$ & \phantom{0}1.7/2 & \phantom{0} $1.43\pm0.09$ \phantom{0} & $2.26\pm0.12$ & 0.24/3 \\
April 26, 2006 & 225 & 62 &                   16.4$\sigma$ & 350 & $1.08\pm0.09$ & \phantom{0}1.3/4 & \phantom{0} $1.21\pm0.11$ \phantom{0} & $2.35\pm0.17$ & 0.41/2 \\
April 27, 2006 & 615 & 56 &                   33.5$\sigma$ & 350 & $3.21\pm0.15$ & \phantom{0}1.9/4 & \phantom{0} $3.37\pm0.18$ \phantom{0} & $2.07\pm0.07$ & \phantom{0}4.8/4 \\
April 28, 2006 & 311 & 75 &                   19.9$\sigma$ & 350 & $1.14\pm0.08$ & \phantom{0}4.3/8 & \phantom{0} $1.32\pm0.10$ \phantom{0} & $2.47\pm0.14$ & 0.65/2 \\
April 29, 2006 & 514 & 169 &                 23.7$\sigma$ & 250 & $1.04\pm0.06$ & 41/7                       & \phantom{0} $1.14\pm0.06$ \phantom{0} & $2.28\pm0.09$ & \phantom{0}2.0/4 \\
April 30, 2006 & 69 &  11 &                    10.3$\sigma$ & 250 & $2.39\pm0.33$ & ---                             & \phantom{0} $2.16\pm0.34$ \phantom{0} & $1.66\pm0.20$ & \phantom{0}1.4/1 \\
June 14, 2006 & 95 & 87 & \phantom{0}7.5$\sigma$ & 450 & $0.34\pm0.06$ & \phantom{0}2.4/1  &                                          $0.168\pm0.032$ & $2.38\pm0.44$ & \phantom{0}1.5/2 \\
\hline
\end{tabular}
\end{center}

{\bf Notes.}
Number of excess ($N_\mathrm{excess}$) and background ($N_\mathrm{backgr.}$)
events, resulting significances $S$, lower cuts in event energy, integral
fluxes $F$ above $E_\mathrm{min}=250\,\mathrm{GeV}$ for the April 2006 data and
$E_\mathrm{min}=450\,\mathrm{GeV}$ for the June 14, 2006 data (in units of
$10^{-10}\,\mathrm{cm}^{-2} \mathrm{s}^{-1}$), fit quality of a constant-flux
fit to the individual observation nights (see Fig.~\ref{fig:lcapril06fine}),
and power-law fit results for the differential energy spectra of
${\mathrm d}F/{\mathrm d}E=f_0 \cdot (E/E_0)^{-\alpha}$ with
$E_0=0.5\,\mathrm{TeV}$ for the April 2006 data and $E_0=1.0\,\mathrm{TeV}$ for
the June 14, 2006 data, respectively; $f_0$ in units of
$10^{-10}\,\mathrm{TeV}^{-1}\mathrm{cm}^{-2} \mathrm{s}^{-1}$.
\end{table*}

The energy thresholds of the individual observations are also given in
Tab.~\ref{tab:aprilspectra}. As the analysis threshold is always lower than the
applied energy cut, the latter one defines the energy threshold value.

The strong $\gamma$-ray signal allowed to infer light curves with a resolution
below one hour for all of the observation nights, which are shown in
Fig.~\ref{fig:lcapril06fine} (see
Tab.~\ref{tab:lightcurvefine} for the
light curve data). Most light curves are compatible with a constant flux during
the nightly observation time (see Tab.~\ref{tab:aprilspectra} for all
constant-fit $\chi^2_\mathrm{red}$ values), while on MJD 53855 a clear
intra-night variability is apparent. A fit with a constant function yields an
unacceptable $\chi^2_\mathrm{red}=41/7$ ($P\approx8\cdot10^{-5}\%$) for this
night, and the data suggest a flux halving time of $36\pm10_{\mathrm stat}$ minutes. Note that
this interesting observation window has also been covered by {\it XMM-Newton}
observations in the X-ray band \citep{acci421}.

\begin{table}
\caption{\label{tab:lightcurvefine} Light curve data.}
\fontsize{9}{9}
\selectfont
\begin{center}
\begin{tabular}{cr}
\hline
Observation      & $F_{>250\,\mathrm{GeV}}$ \\
$[\mathrm{MJD}]$ & $[10^{-10}\mathrm{cm}^{-2}\mathrm{s}^{-1}]$ \\
\hline
\multicolumn{2}{c}{2006/04/22}\\
\hline
53847.98307 & $1.00\pm0.21$\\
53847.99775 & $0.70\pm0.23$\\
53848.00867 & $0.99\pm0.17$\\
\hline
\multicolumn{2}{c}{2006/04/24}\\
\hline
53849.97136 & $2.56\pm0.25$\\
53849.98618 & $2.04\pm0.21$\\
53850.00033 & $2.37\pm0.22$\\
\hline
\multicolumn{2}{c}{2006/04/25}\\
\hline
53850.93996 & $1.24\pm0.15$\\
53850.96431 & $1.49\pm0.15$\\
53850.98652 & $1.26\pm0.15$\\
\hline
\multicolumn{2}{c}{2006/04/26}\\
\hline
53851.93677 & $0.97\pm0.20$\\
53851.95255 & $1.04\pm0.21$\\
53851.96726 & $1.25\pm0.20$\\
53851.98190 & $1.00\pm0.18$\\
53851.99680 & $1.13\pm0.18$\\
\hline
\multicolumn{2}{c}{2006/04/27}\\
\hline
53852.94098 & $3.01\pm0.33$\\
53852.95502 & $3.19\pm0.38$\\
53852.96823 & $3.05\pm0.31$\\
53852.98159 & $3.57\pm0.32$\\
53852.99406 & $3.17\pm0.28$\\
\hline
\multicolumn{2}{c}{2006/04/28}\\
\hline
53853.88754 & $1.36\pm0.25$\\
53853.89880 & $0.80\pm0.27$\\
53853.92893 & $1.22\pm0.24$\\
53853.93984 & $1.09\pm0.25$\\
53853.95457 & $1.18\pm0.24$\\
53853.96887 & $1.22\pm0.27$\\
53853.98040 & $0.95\pm0.19$\\
53853.99316 & $1.32\pm0.21$\\
53854.00687 & $1.12\pm0.18$\\
\hline
\multicolumn{2}{c}{2006/04/29}\\
\hline
53854.90199 & $2.07\pm0.23$\\
53854.91620 & $1.42\pm0.19$\\
53854.95206 & $0.86\pm0.16$\\
53854.96625 & $1.11\pm0.17$\\
53854.97974 & $1.27\pm0.18$\\
53854.99354 & $0.83\pm0.15$\\
53855.00847 & $0.80\pm0.14$\\
53855.02879 & $0.69\pm0.11$\\
\hline
\multicolumn{2}{c}{2006/04/30}\\
\hline
53855.97595 & $2.39\pm0.33$\\
\hline
\multicolumn{2}{c}{2006/06/14}\\
\hline
53900.92797 & $0.45\pm0.09$\\
53900.94585 & $0.26\pm0.08$\\
\hline
\end{tabular}
\end{center}
\end{table}

\subsection{Results for June 14, 2006}
An onset of activity to $\approx 2$ times the average quiescent-flux level of
\object{Mkn\,421} was measured in April 2006 by the {\it RXTE} all-sky monitor
(ASM) instrument. It triggered an {\em INTEGRAL} ToO campaign from June 14, 2006
to 25 for a total of 829~ks \citep{lichtiaa}. This $>30$~mCrab flux remained
until September 2006.  During the 9-day campaign, \object{Mkn\,421} was
targeted by various instruments in the radio, optical, X-ray and VHE wavebands.
Results are reported in \citet{lichtiaa}. On June 14, 2006, MAGIC observed
\object{Mkn\,421} at rather high zenith angles in parallel with the OMC, JEM-X,
and IBIS measurements aboard {\em INTEGRAL}. Further VHE coverage was provided
by the Whipple 10-m telescope on June 18/19/21, 2006
\citep{lichtiaa}.

The MAGIC observations on June 14, 2006 lasted for $\approx$~50 minutes. The
high zenith angles of 43 to 52 degrees of this observations and the previously
mentioned inhomogeneities result in an energy threshold of
$E_\mathrm{thresh.}=450$~GeV. In spite of the overall rather difficult
observational circumstances caused by the high zenith angle observations
\citep{tonello,magicgc}, a firm detection on the $7.5$-$\sigma$ significance
level was achieved. 

\begin{figure}
\begin{center}\includegraphics[width=.8\linewidth]{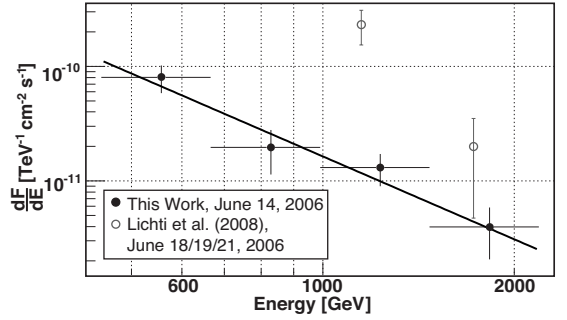}\end{center}
\caption{Differential photon spectrum for \object{Mkn\,421} for the observation
night of June 14, 2006 (black data points). A power-law fit to the spectrum
results in a spectral slope of $\alpha=-2.38\pm0.44$ (See Tab.
\ref{tab:aprilspectra} for the fit results). Also shown are spectral points
measured with the Whipple 10-m telescope \citep{lichtiaa} during June 18-21, 2006.} \label{fig:spectrum15}
\end{figure}

The corresponding differential energy spectrum is shown in
Fig.~\ref{fig:spectrum15}. Between 450~GeV and 2.2~TeV, it can be described by
a simple power-law of the form
\begin{equation}
\frac{\mathrm{d}F}{\mathrm{d}E} = (1.68\pm0.32)\cdot10^{-11}\,
\mathrm{TeV}^{-1}\mathrm{cm}^{-2} \mathrm{s}^{-1} 
\left(\frac{E}{1.0\,\mathrm{TeV}}\right)^{-2.38\pm0.44}
\end{equation}

For comparison we also show the spectral points reported by the Whipple 10-m
telescope averaged over the nights of June 18/19/21, 2006.  Generally,
there might be systematic differences between the Whipple and MAGIC
measurements. It could, however, be shown that such inter-instrument systematic
effects are rather small and under control, e.g.\ those between MAGIC and
H.E.S.S.\ \citep{m05}.  Particularly the \object{Crab nebula} spectra measured
by Whipple and MAGIC agree quite well \citep{magic_cn}. The \object{Mkn\,421}
flux measured by the Whipple 10-m telescope four days after the MAGIC
observation is substantially higher than our measurements (Fig.
\ref{fig:spectrum15}), pointing to a clear evolution of the source emission
level within the {\em INTEGRAL} campaign.

\section{Discussion}\label{discus}
\begin{figure}
\includegraphics[width=\linewidth]{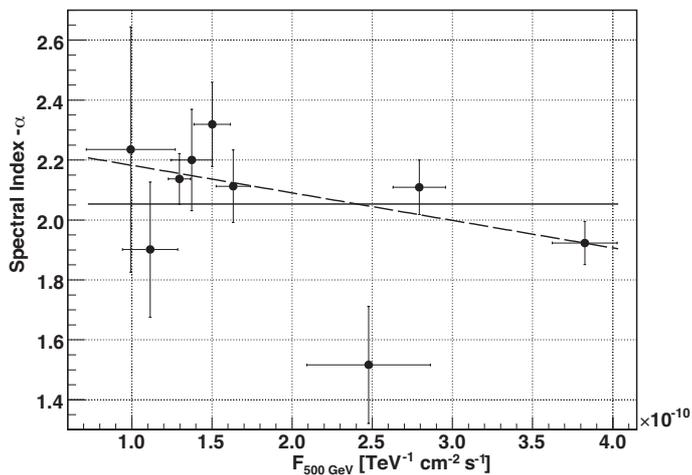}
\caption{Spectral index vs.\ flux at $0.5$~TeV deduced from a simple power-law
fit after EBL de-absorption for \object{Mkn\,421}.
%in April 2006. 
The $\chi^2_\mathrm{red}$ for a constant fit
(spectral index uncorrelated with flux level; solid line) amounts
to $17/8$ ($P\approx3\%$), while a linear correlation (dashed line) has
a $\chi^2_\mathrm{red} = 11/7$, equal to $P\approx12\%$.}
\label{fig:corfluhar}
\end{figure}

In leptonic acceleration models, e.g., SSC models, a shift of the high-energy
peak (attributed to Inverse Compton radiation) in the spectral energy
distribution towards higher energies with an increasing flux level is expected.
In the VHE domain, such a shift can be traced by spectral hardening. Variations
in the hardness of the TeV $\gamma$-ray spectrum during flares were reported by
several groups \cite[e.g.,][]{whipple421,hess421,fossati08}. We tested for a
correlation of the spectral hardness with the flux level of the de-absorbed
spectrum (i.e.\ after removing any attenuation effects caused by the
Extragalactic Background Light [EBL], cf.  \citealt{nikishov,gould,HauserDwek})
in our data (Fig. \ref{fig:corfluhar}), but found that the correlation neither
can be described by a constant fit ($\chi^2_\mathrm{red}=17/8$, $P\approx3\%$)
nor by a linear dependence of spectral hardness and flux level
($\chi^2_\mathrm{red}=11/7$, $P\approx12\%$), giving no clear preference for
either. Although clear flux variations are present in the data set, the overall
dynamical range of $3.9$ in flux might be too small to see a significant
spectral hardening with increasing flux.

\begin{figure*}
\begin{center}
\includegraphics[width=0.8\linewidth]{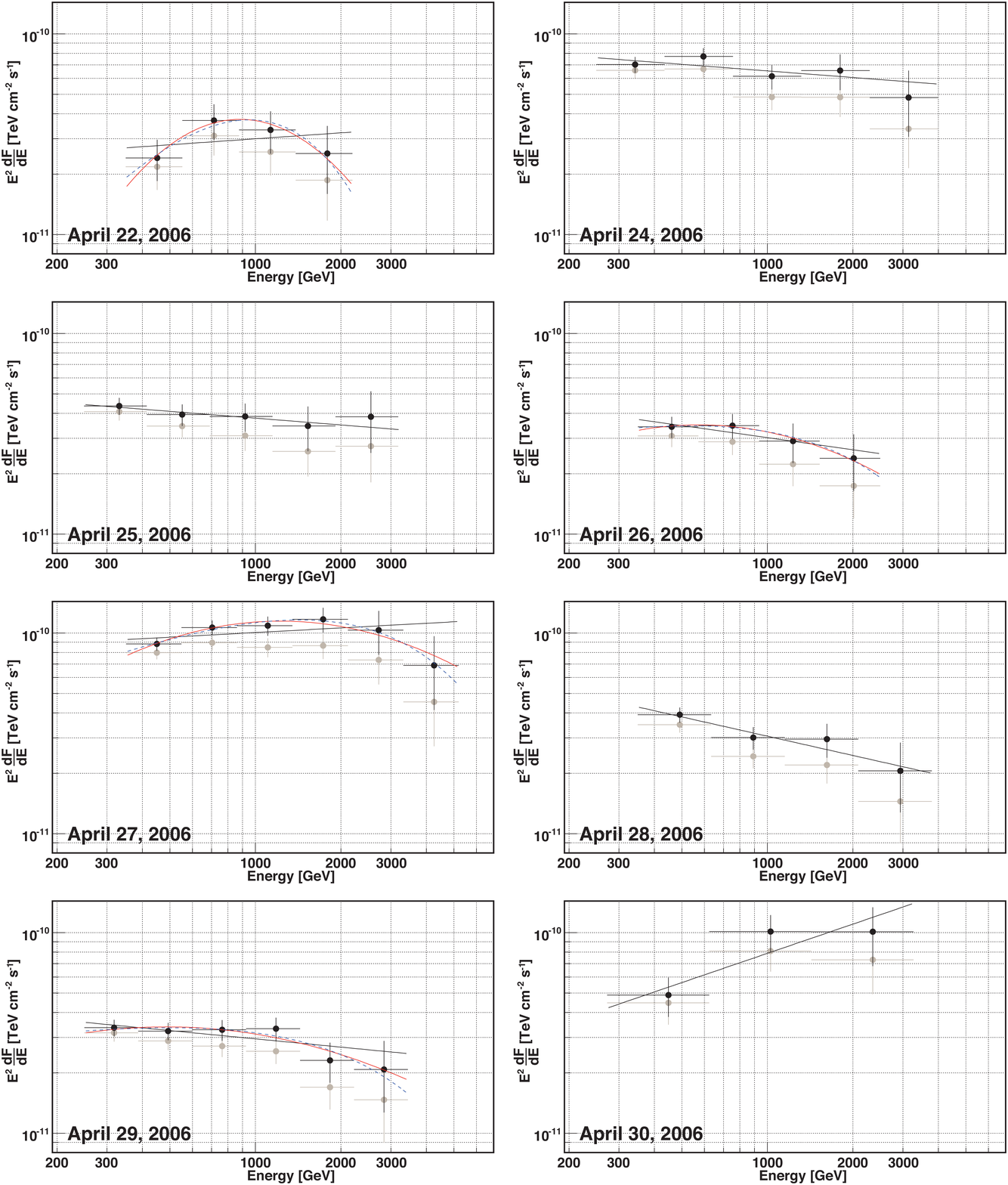}
\end{center}
\caption{Differential energy spectra for \object{Mkn\,421} for
April 2006 before (gray points) and after (black points) correcting for EBL
absorption. For the apparently hard spectra on April 22, 26, 27, and 29, 2006,
log-P (Eq. \ref{eqn:logPL}) and PL+C (Eq.
\ref{eqn:PL+C}) fits were performed (red solid and blue dashed curves, respectively).
} \label{fig:allspec}
\end{figure*}

The individual night-by-night spectra during the campaign in April 2006 are
shown in Fig.~\ref{fig:allspec}. All spectral data points are summarized in
Tab.~\ref{tab:spectra}.
For the nights of April 22, 26, and 29, 2006, there seems to be evidence for a
resolved peak, but a likelihood ratio test  \cite[e.g.,][]{mg07} yields significant curvature only
for April 27, 2006.\footnote{the respective log-P probabilities are 83\%, 48\%, 73\%,
and 96\%.}
We used a logarithmic curvature term, corresponding to
a parabolic power-law (log-P) in a $\log(E^2 \mathrm{d}F/\mathrm{d}E)$ vs.
$\log E$ representation \citep{massaro}, and a power-law with exponential
cutoff (PL+C) of the form
\begin{equation}\label{eqn:logPL}
\frac{\mathrm{d}F}{\mathrm{d}E} = f_0 \cdot  10^{-11}\,
\mathrm{TeV}^{-1}\mathrm{cm}^{-2} \mathrm{s}^{-1}\, 
\left(\frac{E}{E_0}\right)^{-\left(\alpha+\beta \,\log_{10}\left(\frac{E}{E_0}\right)\right)}
\label{eq:4}
\end{equation}
and
\begin{equation}\label{eqn:PL+C}
\frac{\mathrm{d}F}{\mathrm{d}E} = f_0 \cdot  10^{-11}\,
\mathrm{TeV}^{-1}\mathrm{cm}^{-2} \mathrm{s}^{-1}\,
\left(\frac{E}{E_0}\right)^{-\alpha}\,
\exp \left(\frac{-E}{E_{\rm cut}}\right),
\end{equation}
respectively. The likelihood ratio test results in a clear
preference towards a log-P or a PL+C compared to a simple power-law with a
probability of $\approx 96\%$ for both of them. 
The $\chi^2_{\rm red}$ values for 
PL, log-P, and PL+C fits on
the individual night-by-night spectra 
in Fig.~\ref{fig:allspec} are given in
Tab.~\ref{tab:allspecchisq}.
Also the high
statistics data sets defined by combining all data from April, all data from
the five low-state nights and all data from the three high-state nights, clearly
showed evidence for a  parabolic or cutoff shape of the spectra. The results of
the fits and the probability of a likelihood ratio test are given in
Tab.~\ref{tab:curvedspectra}. For all these nights our data did not allow to
prefer one model over the other. The fact that all of the high statistics data
sets show a curved spectral shape is an indication of this feature being always
visible for \object{Mkn\,421} and hence source intrinsic.

\begin{table}
\caption{\label{tab:spectra} Energy spectra for all observation nights under study after EBL de-absorption.}
\begin{center}
\fontsize{9}{9}
\selectfont
\begin{tabular}{rrc}
\hline
\multicolumn{2}{c}{$E$ bounds} & Flux\\
\multicolumn{2}{c}{$[\mathrm{GeV}]$} & $[\mathrm{TeV}\,\mathrm{cm}^{-2}\,\mathrm{s}^{-1}]$\\
\hline
\multicolumn{3}{c}{2006/04/22}\\
\hline
350 &  554 & $(2.39\pm0.56)\cdot10^{-11}$\\
554 &  877 & $(3.67\pm0.75)\cdot10^{-11}$\\
877 & 1389 & $(3.29\pm0.79)\cdot10^{-11}$\\
1389 & 2200 & $(2.51\pm0.93)\cdot10^{-11}$\\
\hline
\multicolumn{3}{c}{2006/04/24}\\
\hline
250 &  435 & $(7.00\pm0.64)\cdot10^{-11}$\\
435 &  758 & $(7.69\pm0.76)\cdot10^{-11}$\\
758 & 1320 & $(6.12\pm0.86)\cdot10^{-11}$\\
1320 & 2297 & $(6.54\pm1.33)\cdot10^{-11}$\\
2297 & 4000 & $(4.80\pm1.74)\cdot10^{-11}$\\
\hline
\multicolumn{3}{c}{2006/04/25}\\
\hline
250 &  416 & $(4.36\pm0.43)\cdot10^{-11}$\\
416 &  693 & $(3.95\pm0.48)\cdot10^{-11}$\\
693 & 1154 & $(3.86\pm0.62)\cdot10^{-11}$\\
1154 & 1922 & $(3.47\pm0.87)\cdot10^{-11}$\\
1922 & 3200 & $(3.85\pm1.31)\cdot10^{-11}$\\
\hline
\multicolumn{3}{c}{2006/04/26}\\
\hline
350 &  572 & $(3.41\pm0.41)\cdot10^{-11}$\\
572 &  935 & $(3.46\pm0.49)\cdot10^{-11}$\\
935 & 1529 & $(2.90\pm0.65)\cdot10^{-11}$\\
1529 & 2500 & $(2.38\pm0.75)\cdot10^{-11}$\\
\hline
\multicolumn{3}{c}{2006/04/27}\\
\hline
350 &  549 & $(8.83\pm0.66)\cdot10^{-11}$\\
549 &  860 & $(1.07\pm0.09)\cdot10^{-10}$\\
860 & 1349 & $(1.09\pm0.12)\cdot10^{-10}$\\
1349 & 2115 & $(1.17\pm0.17)\cdot10^{-10}$\\
2115 & 3317 & $(1.03\pm0.26)\cdot10^{-10}$\\
3317 & 5200 & $(6.89\pm2.76)\cdot10^{-11}$\\
\hline
\multicolumn{3}{c}{2006/04/28}\\
\hline
350 &  635 & $(3.92\pm0.34)\cdot10^{-11}$\\
635 & 1153 & $(3.01\pm0.39)\cdot10^{-11}$\\
1153 & 2093 & $(2.96\pm0.57)\cdot10^{-11}$\\
2093 & 3800 & $(2.06\pm0.79)\cdot10^{-11}$\\
\hline
\multicolumn{3}{c}{2006/04/29}\\
\hline
250 &  387 & $(3.37\pm0.32)\cdot10^{-11}$\\
387 &  600 & $(3.24\pm0.32)\cdot10^{-11}$\\
600 &  929 & $(3.29\pm0.39)\cdot10^{-11}$\\
929 & 1438 & $(3.33\pm0.45)\cdot10^{-11}$\\
1438 & 2228 & $(2.31\pm0.53)\cdot10^{-11}$\\
2228 & 3450 & $(2.08\pm0.81)\cdot10^{-11}$\\
\hline
\multicolumn{3}{c}{2006/04/30}\\
\hline
250 &  572 & $(4.87\pm1.09)\cdot10^{-11}$\\
572 & 1310 & $(1.01\pm0.21)\cdot10^{-10}$\\
1310 & 3000 & $(1.01\pm0.33)\cdot10^{-10}$\\
\hline
\multicolumn{3}{c}{2006/06/14}\\
\hline
450 &  669 & $(2.76\pm0.74)\cdot10^{-11}$\\
669 &  995 & $(1.61\pm0.67)\cdot10^{-11}$\\
995 & 1480 & $(2.54\pm0.79)\cdot10^{-11}$\\
1480 & 2200 & $(1.80\pm0.85)\cdot10^{-11}$\\
\hline
\end{tabular}
\end{center}

{\bf Notes.} The two energy bounds specify the range in which the corresponding flux was measured.
\end{table}

\begin{table}
\caption{\label{tab:allspecchisq} $\chi^2_{\rm red}$ values for the PL, log-P, and PL+C fits performed in Fig.~\ref{fig:allspec}}
%\resizebox{9cm}{!} {
%\begin{tabular}{l|cccccccc}
\begin{tabular}{l|p{.5cm}p{.5cm}p{.5cm}p{.7cm}p{.7cm}p{.5cm}p{.5cm}p{.5cm}}
& 22 & 24 & 25 & 26 & 27 & 28 & 29 & 30 \\
\hline
PL    & 2.2/2  & 1.9/3 & 0.21/3  & 0.47/2 & 5.3/4 & 0.59/2 & 2.3/4 &   1.5/1 \\
log-P & 0.14/1 &  &     & 0.041/1 & 0.48/3 &       & 1.1/3         & \\
PL+C  & 0.27/1 &  &     & 0.076/1 & 0.34/3 &       & 0.87/3        & \\
\hline
\end{tabular}
%}

\vspace{10pt}

{\bf Notes.} The columns represent days in April 2006.
\end{table}

\begin{table*}
\begin{centering}
\caption{\label{tab:curvedspectra} Special fit results.}
\begin{tabular}{llccccccc}
Data Set & Used Fit & $f_0$ & $\alpha^{(')}$ & $\beta$ & $E_{\rm cut}$ [TeV] & $\chi2_\mathrm{red, fit}$ & Likelihood  & $E_{\rm peak}$ [TeV] \\
\hline \hline
\multirow{4}*{April 27, 2006}		& PL      & $9.54\pm0.52$ & $1.92\pm0.07$ & & & \phantom{0}5.3/4 && \\
           & log-P & $9.35\pm0.55$ & $1.54\pm0.19$ & $0.59\pm0.29$ & & 0.48/3 & $96\%$ & \phantom{0} $1.2\pm 0.7$ \phantom{0}\\
							& log-P apex & $11.5\pm0.9\phantom{0}$ & $0.26\pm0.17$ & &  & 0.48/3 & $96\%$ & $1.2\pm0.2$ \\
							& PL+C & $11.3\pm1.2 \phantom{0}$ & $1.44\pm0.24$ && $2.6\pm1.3$ & 0.34/3 & $96\%$ & \phantom{0} $1.4\pm 1.0$ \phantom{0} \\\hline
\multirow{4}*{All April Data}		& PL      & $4.53\pm0.07$ & $2.07\pm0.04$ & & & \phantom{0} 16/5&& \\
           & log-P & $4.75\pm0.12$ & $1.89\pm0.06$ & $0.39\pm0.11$ & & \phantom{0}1.2/4 & $99\%$ & $0.69\pm 0.14$ \\
							& log-P apex &  $4.84\pm0.16$ & $0.41\pm0.11$ & & & \phantom{0}1.2/4 & $99\%$ & $0.69\pm0.06$ \\
							& PL+C &  $5.36\pm0.31$ & $1.77\pm0.09$ & & $3.6\pm1.1$ & \phantom{0}1.8/4 & $99\%$&  \phantom{0} $0.80\pm 0.42$  \phantom{0} \\\hline
\multirow{4}*{High-State Nights}	& PL      &  $8.19\pm0.28$ & $1.93\pm0.05$ & & & \phantom{0}6.0/4 && \\
           & log-P &  $8.46\pm0.32$ & $1.79\pm0.09$ & $0.29\pm0.15$ & & \phantom{0}2.0/3 & $94\%$ & \phantom{0} $1.1\pm 0.6$ \phantom{0} \\
							& log-P apex &  $9.21\pm0.47$ & $0.52\pm0.17$ & &  & \phantom{0}2.0/3 & $94\%$ &  $1.1\pm0.3$ \\
							& PL+C &  $9.02\pm0.64$ & $1.75\pm0.12$ & & $6.1\pm4.0$ & \phantom{0}3.1/3 & $91\%$& \phantom{0} $1.5\pm 1.2$ \phantom{0}\\\hline
\multirow{4}*{Low-State Nights}	& PL      &  $3.39\pm0.10$ & $2.17\pm0.05$ & & & \phantom{0}6.6/4 && \\
           & log-P &  $3.55\pm0.13$ & $2.02\pm0.08$ & $0.38\pm0.17$ & & \phantom{0}1.1/3 & $97\%$ & $0.48\pm 0.12$ \\
							& log-P apex &  $3.55\pm0.13$ & $0.41\pm0.16$ & & & \phantom{0}1.1/3 & $97\%$&  $0.48\pm0.12$  \\
							& PL+C &  $4.15\pm0.40$ & $1.85\pm0.15$ & & $2.9\pm1.3$  & 0.75/3 & $97\%$ & $0.45\pm 0.47$ \\
\hline
\end{tabular}

{\bf Notes.} Results of a power-law fit (Eq.\ \ref{eq:2}), a log-parabolic fit (Eq.\ \ref{eq:4}  $\cdot
(E/E_0)^2$), a log-parabolic fit in apex form (Eq.\ \ref{eq:apex}) and
a power-law fit with an exponential cutoff ((Eq.\ \ref{eqn:PL+C}) $\cdot
(E/E_0)^2$) in $E^2 \mathrm{d}F/\mathrm{d}E$ after EBL de-absorption for
special data sets. $f_0$ is given in units of
$10^{-11}\mathrm{TeV}^{-1}\mathrm{cm}^{-2} \mathrm{s}^{-1}$; $\alpha$, $\alpha'$, $\beta$, $E_{\rm cut}$ and
$E_{\rm peak}$ are the fit parameters as stated in the text, and Likelihood denotes
the probability of a likelihood ratio test. The on/off normalization factor is $1/3$, $E_0=0.5$~TeV.
\end{centering}
\end{table*}

\begin{figure}
\begin{center}\includegraphics[width=.8\linewidth]{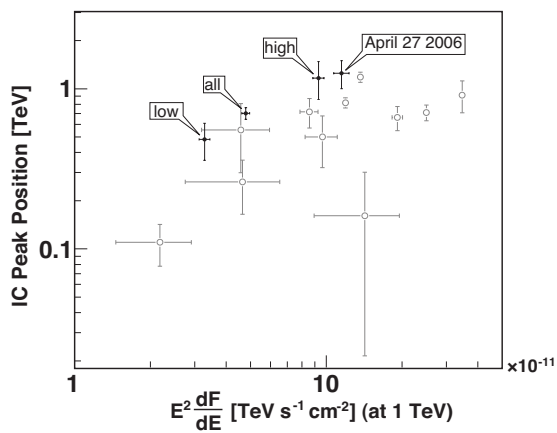}\end{center}
\vspace*{-.6cm}
\caption{Derived peak position using the log-P (Eq.\ \ref{eq:apex}) versus
flux at 1~TeV for the data sets presented in Tab.~\ref{tab:curvedspectra}.
Historical data, taken from \citet{magic421}, are shown in gray. Our data
confirm the indication of a correlation between the two parameters.}
\label{fig:peakvsflux}
\end{figure}

The curved power laws enable to locate a peak in the de-absorbed spectrum at
$E_{\rm peak} = E_0  10^{(2-\alpha)/(2 \beta)}$ for the log-P and at
$E_{\rm peak} = (2-\alpha) E_{\rm cut}$ if $\alpha < 2$ for the PL+C fit.
For simplicity we
determined $E_{\rm peak}$ of the log-P by using the apex form of the parabola in a
logarithmic representation:
\begin{equation}
\log_{10} \frac{{\rm d}F}{{\rm d}E} = \log_{10} f_0 +
\log_{10} \alpha' \left(\log_{10}\left(\frac{E}{E_0}\!\mathbin{\bigg/}\!\frac{E_{\rm peak}}{E_0}\right)\right)^2 \label{eq:apex}
\end{equation}
which naturally yields both  $E_{\rm peak}$ and the flux at the peak, $f_0$,
respectively.  Additionally, the spectral cutoff is naturally obtained from the
PL+C fit as the fit parameter $E_{\rm cut}$. The results are shown in
Tab.~\ref{tab:curvedspectra}. The values of $E_{\rm peak}$ as determined using
the log-P and the PL+C were compatible with each other for the data sets
averaging several nights and showed indications for an increase of the peak
energy with rising flux level, as predicted if the VHE radiation were due to
SSC mechanisms.  We compare our results with historical values taken from
\citet{magic421} in Fig.~\ref{fig:peakvsflux}. Our data
confirm the previously suggested correlation.

The observation of a relation between flux (and thus, fluence) and the position
of the VHE peak in the SED could be signalling a relation similar to the one
suggested by \citet{amati} and observed by \citet{sakamoto} for gamma-ray
bursts. Since the TeV $\gamma$-ray production is assumed to take place in a
relativistic jet, and many of the same radiative processes are involved (on a
larger scale, of course) it might be a similar (or related) mechanism at work
on a different scale. A trend towards a relation between flux and spectral
index in the TeV energy range has also been noted by \citet{wagner08a},
studying 17 known TeV blazars, and by \citet{Tramacere2010} in the X-ray band,
after a deep spectral analysis of all {\it Swift} observations of \object{Mrk 421}
between April and July 2006.

Although the peak energy measured on April 27, 2006 exceeds that of the All
April Data and Low-State data set, it is, despite having a higher flux,
comparable with that derived for the High-State data set. This discrepancy in
terms of the expected behaviour in SSC models can be explained with the
different nature of the data sets: The April 27, 2006 data represent a rather
particular, $1.4$~h long episode of an individual flare event, whereas the
High-State data set is an average of three individual flares. Due to the sparse
sampling, most probably each of these observations caught different epochs of
the individual flare evolutions, during which the spectral shape can change
considerably in terms of spectral index and curvature \cite[see,
e.g.,][]{katarzynski}. Hence the two data sets are not necessarily directly
comparable.

The values of the derived cutoff energies are also suggesting this behavior,
showing, with the exception of April 27, 2006, an increase with rising flux, thus
indicating a source-intrinsic rather than a cosmological reason for the cutoff
feature. This is in accordance with the \citet{kneiskelow} lower-limit model,
predicting an EBL cutoff for \object{Mkn\,421} at around 13~TeV. 

\begin{figure}
\includegraphics[width=\linewidth]{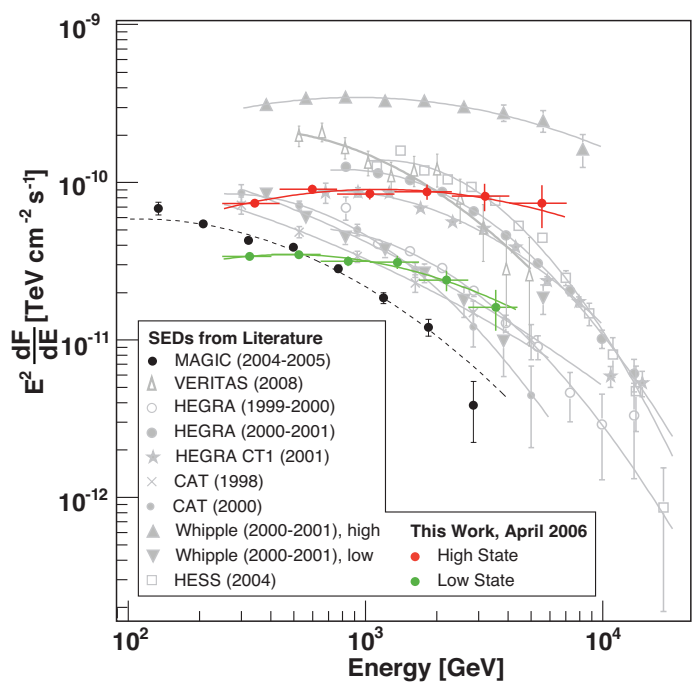}
\caption{EBL de-absorbed historical spectra of \object{Mkn\,421} \cite[see][for
references]{magic421} along with selected spectra from the April 2006 campaign
and the flare spectrum of \citet{donna09}. The solid line is the result of a
fit using Eq.\ \ref{eqn:logPL}. 
Note that the historical data were deabsorbed using the model of
\citet{primack05}, our data and those from \citet{donna09} with the model of
\citet{kneiskelow}.}
\label{fig:hist}
\end{figure}

In Fig.~\ref{fig:hist}, we compare ``historical'' spectra measured between 1998
and 2005 with the low-state and high-state spectra derived from the
observations reported here. It is obvious that our low-state spectrum
represents one of the lowest flux states ever measured in VHE for
\object{Mkn\,421}, whereas the high state spectrum shows no exceptionally high
flux level of this source.  Both spectra are harder than historical spectra
with comparable flux levels, in particular harder than the VERITAS spectrum
\citep{donna09}, enabling one of the best measurements of the turnover of the
SED in a low flux state.  While a previous observation yielded a rather flat
spectrum in the VHE regime \citep[][]{hegra421}, we conclude that we measured a
rather clear peak (flat structure in the SED).  The low-state spectrum has a
shape similar to the one measured by HEGRA CT1, although at an approximately
three times lower flux level.  The high-state spectral shape resembles the
high-state Whipple spectrum, which in turn has an about three times higher
flux.  This tendency can also be seen in Fig.~\ref{fig:peakvsflux}, which shows
that the fluxes we derive are systematically lower than historical measurements
for comparable peak energies.  Within the SSC framework this difference in flux
for comparable spectral shapes can be caused by, e.g., a lower number of
electrons with the same energy distribution as in the high-flux case.

In summary, we followed the evolution of a sequence of mild flares of the
blazar \object{Mkn\,421} during one week from April 22 to 30, 2006, peaking at
$F(E>250\,\mathrm{GeV})=
(3.21\pm0.15)\,10^{-10}\,\mathrm{cm}^{-2}\mathrm{s}^{-1}$ ($\approx 2.0$ Crab
units). The nocturnal observations lasted at least for about one hour and allowed
for the reconstruction of night-by-night spectra. During three observation
nights high fluxes were recorded, in which, however, no variability could be
measured. In two of these nights, rather hard spectral indices were found, but
this was also the case for the night with the lowest flux. During the night of
April 29, 2006, with a not particularly high flux of $F(E>250\,\mathrm{GeV})=
(1.04\pm0.06)\,10^{-10}\,\mathrm{cm}^{-2}\mathrm{s}^{-1}$ ($\approx 0.65$ Crab
units), clear intra-night variability with a flux-doubling time of $36\pm10_{\mathrm stat}$
minutes was observed.

According to a likelihood ratio test, the spectra of some data sets were better
described by curved power laws than simple power laws, enabling us to calculate
peak and cutoff energies in the VHE regime. The derived peak values are
consistent with an evolution of the peak energy with the flux, as suggested by
historical data. Indications of an intrinsic cutoff in the spectra of Mkn 421,
as found in former observations, are confirmed by our results.

During the {\em INTEGRAL}-triggered MWL campaign in June 2006 we observed
\object{Mkn\,421} in one night at high zenith angles. Our measurements
complement the three-night observations conducted by the Whipple 10-m telescope
four days later. Taking the MAGIC and Whipple results together, a variability
of \object{Mkn\,421} also during the {\em INTEGRAL} observations is evident.
The energy coverage of the Whipple telescope spectrum ($\Delta~E~\approx
600$~GeV) was not sufficient to assess any spectral evolution by comparing it
to the MAGIC spectrum ($\Delta~E~\approx 2$~TeV).

The determined fluxes and spectra will be further used for studies of the SED
taking into account data taken at other photon energies in detailed MWL
analyses (publications in preparation).

\acknowledgements
We thank the Instituto de Astrof\'{\i}sica de Canarias for the excellent
working conditions at the Observatorio del Roque de los Muchachos in La Palma.
The support of the German BMBF and MPG, the Italian INFN and Spanish MICINN is
gratefully acknowledged.  This work was also supported by ETH Research Grant TH
34/043, by the Polish MNiSzW Grant N N203 390834, and by the YIP of the
Helmholtz Gemeinschaft.

\end{document}